\documentstyle[preprint,aps]{revtex}
\def\upss{\Upsilon (3S) \rightarrow \Upsilon (1S) \ \pi \pi}

\begin{document}
% \draft command makes pacs numbers print
\draft
\preprint{TPI--MINN--93/49--T
\hspace{-38.0mm}
\raisebox{2.4ex}{UMN--TH--1207/93}}
%\tighten
\title{Final--state $\pi \pi$ interactions \\
in $\Upsilon (3S) \rightarrow \Upsilon (1S) \ \pi \pi$ }
\author{Sumantra Chakravarty,  Sun Myong Kim    \\
and \\
Pyungwon Ko
%\thanks{ pyungwon@mnhep.hep.umn.edu}
\\ }
\address{ School of Physics and Astronomy \\ University of Minnesota \\
Minneapolis, MN 55455 \\ }
%\date{\today}
\maketitle
\begin{abstract}
The $m_{\pi\pi}$ spectrum and various  angular distributions in $\upss$ are
studied including the effects of  the $\pi\pi$ phase shift in the $I=L=0$
channel using the lowest order amplitude in the pion momentum expansion.
Our results are compared with the recent CLEO data, and we find good
agreement except for the $\cos\theta_{\pi}^*$ distributions.
We argue that the $\cos\theta_{\pi}^*$ distribution, contrary to other
distributions,  is sensitive to  the higher order corrections in the
pion momentum expansion.    This argument is supported by using an ansatz
for the amplitude which is of higher order in the pion momentum expansion
and still satisfies the soft pion theorem.
\\
\end{abstract}

% insert suggested PACS numbers in braces on next line

% body of paper here

\narrowtext
%\tighten

\section{Introduction}
\label{sec:intro}
%{\it 1. Introduction}

The peculiar double peaks in the $m_{\pi \pi}$ spectrum in $\upss$ \cite{cusb}
have   been lacking proper understanding.  Although several suggestions have
been  made, most of them reproduce the $\pi \pi$ spectrum only with limited
success \cite{prework}.
In the two recent works by us, we approached this problem in two different
ways.  In the first work \cite{chakrako}, we assumed that the most general
form of the amplitude for $\upss$  is
\begin{equation}
{\cal M} = A_{0} \left[ \left( q^{2} + B_{0} E_1 E_2 + C_{0} m_{\pi}^2
\right)~\epsilon \cdot  \epsilon^{'} + D_{0} \left( p \cdot \epsilon
p^{'} \cdot \epsilon^{'} + p \cdot \epsilon^{'} ~p^{'} \cdot
\epsilon \right) \right],
\label{eq:amp1}
\end{equation}
in the lowest order in the pion momentum expansion.
Here, $p, p^{'}$ are the four--momenta of the final pions, $E_1$ and
$E_2$ are their energies, and $\epsilon$ and $\epsilon^{'}$ are the
polarization vectors of the initial and the final $\Upsilon$'s,
respectively.  In Ref.~\cite{chakrako}, we found three sets of
parameters (P0, P1 and P2)  by minimizing the $\chi^2$.
For these three sets of parameters, we predicted various angular
distributions which should be checked against experiments.

In the second work \cite{chakraetal},  we assumed that (i) the QCD
multipole expansion is applicable to $\upss$ and (ii) $\Upsilon (3S)$ has
an admixture of a $D-$wave component :
\begin{equation}
| \Upsilon (3S) \rangle = \cos \phi |3S \rangle  + \sin \phi | D \rangle.
\label{eq:mixing}
\end{equation}
The amplitude for $\upss$ then
depends on three independent parameters, and we found two fits
which correspond to P1 and P2 of Ref.~\cite{chakrako}, respectively.
We then explored the consequences of our assumptions on other
hadronic and radiative transitions of $\Upsilon (3S)$ into the lower
level bottomonia.
In Ref.~\cite{chakraetal}, we again assumed that the lowest order
expansion in the pion momenta
\begin{equation}
\theta_{2 \pi}^{0} (q^2) \equiv \langle \pi \pi | \theta_{\mu}^{\mu}
| 0 \rangle = ( q^2 + m_{\pi}^2 )
\label{eq:theta1}
\end{equation}
be valid  through the whole range of $m_{\pi \pi} = \sqrt{q^{2}}$.

Although these two approaches fit the $m_{\pi\pi}$ spectrum,
they still leave room for theoretical improvement in two aspects.
First of all,  it is well known from the analysis of the $\pi \pi$ phase
shift that the dipion system in $I = L = 0$ experiences strong final
state interactions \cite{truong}.  Since the dipion system
in $\upss$ are in either $I = L = 0$ or $I = 0, L = 2$ state,  one should
properly take into account of the $\pi\pi$ phase shift due to the
final state interactions in the $I = L = 0$ dipion system.

Secondly, the validity of (\ref{eq:amp1}) or  (\ref{eq:theta1}) in $\upss$
is rather unclear, since the available $m_{\pi \pi}$ in $\upss$ is  large,
$ 2 m_{\pi} \leq m_{\pi \pi} \leq ( m_{i} - m_{f} ) = 895$ MeV.
The amplitude (\ref{eq:amp1}) with $B_0 = D_0 = 0$ gives a good description
for the $m_{\pi\pi}$ spectrum in $\Upsilon (2S) \rightarrow \Upsilon(1S)
\ \pi\pi$, where $m_{\pi\pi} \leq 563$ MeV.    For higher value of
$m_{\pi\pi}$,   we simply assume that it is valid and explore its consequences
on various  spectra of decay products in $\upss$.
If any of the predicted angular distributions based on the amplitude
(\ref{eq:amp1}) including the $\pi\pi$ phase shift does not agree with
the experimental measurements, it would signal the importance of higher
order terms in the pion momenta which have been neglected in
(\ref{eq:amp1}).

In the present work, we follow the approach of Ref.~\cite{chakrako} and
include the phase shift of the $\pi\pi$ system in the $I=L=0$ channel
($\delta_{0} (q^{2})$)   using the data
available in the literature \cite{pshift}.
The phase shift for the $I = 0$ $D-$wave $\pi\pi$ system ($\delta_{2}
(q^{2})$) is tiny enough to be neglected for the whole range of
$m_{\pi\pi}$ \cite{pshift}.

In Sec.~\ref{subsec:predict}, we decompose the amplitude (\ref{eq:amp1})
into the $\pi\pi$ $S-$ and $D-$waves, and incorporate the phase shift
$\delta_{0} (q^{2})$.  It is found that
the importance of the phase shift due to the final state interactions is most
prominent in the $\cos \theta_{\pi}^*$  distributions, but the effect is
only moderate.   Then, our  results are compared with the recent data
from CLEO in  Sec.~\ref{subsec:data}.
In Sec.~\ref{sec:three},  we generalize the amplitude (\ref{eq:amp1})
(in case of $D = 0$)  to
include higher order terms in $q^2$, and discuss some general aspects of the
various angular distributions.  It is also pointed out that one can extract
the difference between $\pi\pi$ phase shifts of the $S-$ and $D-$waves
in the $I=0$ channel, $\delta_{0} - \delta_{2}$,
by  measuring the joint distribution ${\rm d}^{2}\Gamma / {\rm d}m_{\pi\pi}
{\rm d} \cos \theta_{\pi}^*$.
The results are summarized in Sec.~\ref{sec:four}.

\section{ effects  of the $\pi\pi$ phase shift}
\label{sec:two}

\subsection{ Predictions on the $\cos \theta_{\pi^*}$ distributions}
\label{subsec:predict}

In this work, we use a modified form of the amplitude (\ref{eq:amp1}).
First of all, we ignore the recoil of the final $\Upsilon (1S)$ in the
amplitude, and make the $D-$term proportinal to the symmetric traceless
part :
\begin{equation}
{\cal M} = A~\left[ \left\{ q^{2} + B E_{1} E_{2} + C m_{\pi}^{2}
\right\} \hat{\epsilon} \cdot \hat{\epsilon}^{'} + D \left\{
\vec{p} \cdot \hat{\epsilon} \vec{p}^{'} \cdot \hat{\epsilon}^{'}
+ \vec{p}^{'} \cdot \hat{\epsilon} \vec{p} \cdot \hat{\epsilon}^{'}
 - {2\over 3} \vec{p} \cdot \vec{p}^{'}  \hat{\epsilon} \cdot
\hat{\epsilon}^{'} \right\} ~\right].
\label{eq:ampnew1}
\end{equation}
One can find relations between parameters in  (\ref{eq:amp1}) and
(\ref{eq:ampnew1}) using
\[
\vec{p} \cdot \vec{p}^{'} = E_{1} E_{2} - {1\over 2}~(s - 2 m_{\pi}^{2}).
\]
It should be emphasized  that our amplitudes  (\ref{eq:amp1}) and
(\ref{eq:ampnew1})  satisfy  Adler's condition by construction.

Now, we decompose the above amplitude into the $\pi\pi$ $S-$wave and
$D-$wave in order to take into account the phase shift of the $\pi\pi$
system in the $I = L = 0$ state.    The $S-$ and $D-$waves have  the
following tensor structures \cite{belanger} :
\begin{eqnarray}
{\cal S}_{ij} & = & f_{S}(q^2) \delta_{ij} + g_{S}(q^2) ( q_{i} q_{j} +
q^{2} \delta_{ij} ), \cr
{\cal D}_{ij} & = & f_{D}(q^2) \left( \cos^{2}\theta_{\pi}^{*} -
{1\over 3} \right)~
\delta_{ij} + g_{D}(q^2) \left[ r_{i} r_{j} -
{1\over 3} (  q_{i} q_{j} + q^{2} \delta_{ij} ) \beta_{\pi}^{*2} \right],
\label{sdij}
\end{eqnarray}
where $r_{\mu} = p_{\mu} - p_{\mu}^{'}$.
The amplitude (\ref{eq:ampnew1}) can be decomposed into the  $f_{S,D},
g_{S,D}$ form factors as follows :
\begin{eqnarray}
f_{S} & = & q^{2} + C m_{\pi}^{2} + {B\over 4}~\left[ (E_{1} + E_{2})^{2}
- {1\over 3}~\vec{q}^{2} \beta_{\pi}^{*2} \right]
\nonumber    \\
& & - {1\over 2}~D q^{2} -{1\over 6}~D~\left[ \vec{q}^{2} -
\beta_{\pi}^{*2}\left( q^{2}
+ {1\over 3} \vec{q}^{2} \right)  \right],
\label{eq:fs}
\\
g_{S} & = & {1\over 2}~D~( 1 - {1\over  3} \beta_{\pi}^{*2} ),
\label{eq:gs}
\\
f_{D} & = & \left( {1\over 6} D - {1\over 4} B \right)~\beta_{\pi}^{*2}~
\vec{q}^{2},
\label{eq:fd}
\\
g_{D} & = & -{1\over 2}~D.
\label{eq:gd}
\end{eqnarray}
Multiplying the $S-$wave amplitude by the phase shift $\delta_{0}
(q^{2})$, we get
\begin{equation}
{\cal M} = \left[ {\cal S}_{ij} e^{i \delta_{0} (q^{2})} +
{\cal D}_{ij} \right]~\hat{\epsilon}_{i} \hat{\epsilon}_{j}^{'}.
\label{eq:ampnew}
\end{equation}

In case the final $\Upsilon (1S)$ is not reconstructed, summations
over polarizations of the initial and final $\Upsilon$'s are done with
\begin{eqnarray}
\sum \hat{\epsilon}_{i} \hat{\epsilon}_{j} & = & \left( \delta_{ij} -
\hat{z}_{i} \hat{z}_{j} \right),   \label{eq:pol1}  \\
\sum \hat{\epsilon}_{i}^{'} \hat{\epsilon}_{j}^{'} & = & \delta_{ij}.
\label{eq:pol2}
\end{eqnarray}
The fact that the initial $\Upsilon (3S)$ is transversely polarized
with respect to the beam directions (taken along the $z-$direction)
has been taken into account in (\ref{eq:pol1}).
This fact is very  useful to test the existence of the $D-$term by
measuring the polar  angle distributions of the final $\Upsilon (1S)$
(or, equivalently, the $\pi\pi$ system as a whole) and/or of a muon emerging
from the  muonic  decay of the  final $\Upsilon (1S)$.
If one tags the muonic decay of the final $\Upsilon (1S)$, the polarization
sum over the final $\Upsilon (1S)$ (\ref{eq:pol2}) should be replaced by
\begin{equation}
\Sigma \hat{\epsilon}_{i}^{'} \hat{\epsilon}_{j}^{'} = ( \delta_{ij} -
\hat{l}_{i} \hat{l}_{j} ),
\end{equation}
where $\hat{l}$ is the three dimensional  unit vector   along the
direction of a muon  in the rest frame of the initial $\Upsilon (3S)$.
For $D=0$, one gets
\[
dN/d\cos \theta_{l} \sim (1+\cos^{2} \theta_l),
\]
where $\cos\theta_{l} = \hat{l} \cdot \hat{z}$.

Using amplitude (\ref{eq:ampnew}), we fit the $m_{\pi\pi}$ spectrum
by minimizing $\chi^2$.
The best fit is given by three sets of solutions, P0, P1 and P2
(see Table~1), which are essentially the same as the ones given in Ref.~
\cite{chakrako}.   Other angular distributions can be obtained by numerical
integrations as in \cite{chakrako}.
In Fig.~\ref{figone} (a) and (b), we show the $\cos \theta_{\pi}^*$
distributions of $\pi^+$ in the rest frame of the dipion system for P0
and P1 (P2), where $\theta_{\pi}^{*} = 0^{\circ}$
is along the direction of the dipion system as a whole in the rest
frame of initial   $\Upsilon (3S)$.
For comparison, we show the corresponding plots with the phase shift
neglected ({\it i.e.} $\delta_{0} (q^{2}) = 0$)   in Fig.~2 (a) and (b).
The phase shift moderately changes the  $\cos \theta_{\pi}^*$ distributions,
but the overall effects may  be hardly discernible in the experiment.

Before continuing on to the next section, we discuss how much the results
obtained in Ref.~\cite{chakraetal} will change when we incorporate the
$\pi\pi$ phase shift and  possible corrections to (3) in higher orders
in the pion momentum expansion.   In this case, it suffices  to resort to
the QCD multipole expansion by our assumptions.   The relevant matrix
element for $\langle \pi \pi | G_{\mu \nu}^{a} G^{a\mu\nu} | 0 \rangle$
has been obtained by Donoghue {\it et al.} using the dispersive approach in
conjunction with the chiral symmetry relations imposed in the chiral
limit \cite{gasser}.   The result is that the above matrix element,
$\langle \pi \pi | G_{\mu \nu}^{a} G^{a\mu\nu} | 0 \rangle$, is
dominated by $\langle \pi\pi | \theta_{\mu}^{\mu} | 0 \rangle$, and that
(3) remains essentially unchanged up to $m_{\pi\pi} \sim 0.9$ GeV,
once it is regarded as the modulus of $\theta_{2\pi}$.
The phase shift is given by $\delta_{0} (q^{2})$.  In short, one only
has to write (3) as
\begin{equation}
\langle \pi\pi | \theta_{\mu}^{\mu} | 0 \rangle = ( q^{2} + m_{\pi}^{2} ) ~
e^{i \delta_{0} (q^{2})}.
\end{equation}
Since the phase shift $\delta_{0} (q^{2})$ does not affect the $m_{\pi\pi}$
spectrum in $\upss$, our results derived in Ref.~\cite{chakraetal} do not
change at all. In particular, the predictions for  $\Upsilon (3S) \rightarrow
\Upsilon (1S) + \eta$ remain  the same, which  excludes the fit P1
\cite{chakraetal}.

\subsection{Comparisons with the data}
\label{subsec:data}

Recently, the CLEO collaboration released a new set of data on hadronic
transitions in $\Upsilon (3S)$ decays \cite{newcleo}.
Their results on $\upss$ can be  summarized as follows :

\vspace{.2in}

(i) the $\cos \theta_{f}$ distribution is flat for the whole range of $
\cos\theta_f$.

(ii) the $\cos\theta_{\pi^*}$ distributions can be fitted by $(a + b \cos
\theta_{\pi^*}^{2})$, with $a = (1.24 \pm 0.06)$

{}~~~and $b = (-0.49 \pm 0.13)$.

(iii) the $\cos \theta_l$ distribution is  consistent with $( 1 + \cos
\theta_{l}^{2} )$ for $ 0 < | \cos\theta_{l} | < 0.7$.

\vspace{.2in}

\noindent
Let us discuss the implication of  each statement above to our fits, P0--P2.
Statement (i) excludes both P1 and P2, since these two lead to quadratic
functions of $\cos\theta_f$.  Statement (iii) also partly supports this
conclusion, since (iii) implies $D=0$  in the amplitude (1).
Thus, (i) and (iii) select P0 as the final candidate.
However, the $\cos\theta_{\pi^*}$ distribution shown in Fig.~\ref{figone}
(a) does not agree with statement (ii) from CLEO.
We do not interpret this disagreement of the CLEO data with our
prediction on the $\cos \theta_{\pi}^*$ distribution
as a general failure of our approach  based on the matrix  element
satisfying the soft pion theorem.  We rather regard
it as  an indication  that amplitude (1) needs  to be modified to include
higher order terms in the pion momentum expansion.
This will be  illustrated  in the next section with a modified
amplitude for $\upss$.

\section{more on the amplitude with $D=0$}
\label{sec:three}

In this section,   we consider the case $D=0$ in more detail, including
possible higher order terms in $q^2$ in the $S-$wave $\pi\pi$ amplitude,
$f_{S}$.  It will be shown that the $\cos \theta_{\pi}^*$
distribution is  sensitive to such  higher order terms  in $q^2$
contrary to other distributions.

Let us write the amplitude for $\upss$ as
\begin{equation}
{\cal M} (\upss) = A~\left[~ f_{S} (q^{2}) e^{i \delta_{0} (q^{2})} +
f_{D} (q^2) ( \cos^{2} \theta_{\pi}^{*} - {1\over 3} ) ~\right]~\hat{
\epsilon}
\cdot \hat{\epsilon}^{'},
\label{eq:ampfin}
\end{equation}
where $f_S$ and $f_D$ satisfy the soft pion theorem.
The explicit forms of $f_{S,D}$ for the lowest order amplitude
(\ref{eq:ampnew1}) can be read off from  (\ref{eq:fs})--(\ref{eq:gd})
with $D=0$.  Therefore the differential cross section for $ e^+ e^-
\rightarrow \upss \rightarrow \pi\pi\mu^{+}\mu^{-} $ is given by
\begin{eqnarray}
%\FL
d^{3} \Gamma & \propto  &
dm_{\pi\pi} d\cos\theta_{\pi}^{*} d\cos\theta_{l} ~~
m_{\pi\pi} | \vec{q} |~\beta_{\pi}^{*}~\left[ 1 + \cos^{2}
\theta_{l}  \right]    \nonumber    \\
&  \times &
\left[ ~f_{S}^{2} + f_{D}^{2} \left( \cos^{2} \theta_{\pi}^{*} - { 1 \over
3} \right)^{2} + 2 f_{S} f_{D} \cos\delta \left( \cos^{2} \theta_{\pi}^{*}
- { 1\over 3} \right) ~\right],
\label{eq:dcrosect}
\end{eqnarray}
where $\beta_{\pi}^*$ is the velocity
of a pion in the $\pi\pi$ rest frame and  $\theta_{l}$ is the
angle between a muon and the $e^+ e^-$ beam in the rest frame of $\Upsilon
(3S)$.

Integrating the partial distribution (\ref{eq:dcrosect})  over
appropriate   variables, one gets
\begin{eqnarray}
d\Gamma\over d\cos\theta_{f} & \propto & 1, ~~~~({\rm flat \ \ distribution}),
\label{eq:cosf}   \\
d\Gamma\over d\cos\theta_{l} & \propto & ( 1 + \cos^{2} \theta_{l} ).
\label{eq:cosmu}
\end{eqnarray}
These two distributions are independent of the $q^2$ dependence
of the form factors, $f_{S}(q^{2})$ and $f_{D}(q^{2})$, as well as of
the $\pi\pi$ phase shift.
And, these results are  consistent with the recent report from CLEO.

On the other hand,  the $\cos\theta_{\pi}^*$ distribution is sensitive
to the actual forms of $f_{S}(q^{2})$ and $f_{D}(q^{2})$, and to the
$\pi\pi$ phase shift.
In principle, there are many possible terms to the next order in the
pion   momentum expansion.  Instead of writing down all possible
terms and fitting the $m_{\pi\pi}$ spectrum as in Ref.~\cite{chakrako},
we  take  the following amplitude  for illustration :
\begin{equation}
f_{S} (q^{2}) = q^{2} \left[ 1 + C \left( {E_{1} + E_{2} \over
m_{\pi}}  \right) \right]
+ {B\over 4}~\left[
( E_{1} + E_{2} )^{2} - {1\over 3} |\vec{p}_{f}|^{2} \beta_{\pi}^{* 2}
\right],
\label{eq:fsex}
\end{equation}
with the same  $f_{D} (q^{2})$  as before.  This amplitude has three
parameters, $A,B$ and $C$, and satisfies the soft pion theorem like
(\ref{eq:amp1}).

By $\chi^{2}$ fit to the $m_{\pi\pi}$ spectrum,
we found another fit (we will call it P3) with $\chi^{2}/d.o.f. = 11.0/7$ (see
Fig.~ \ref{figthree} (a)).  The corresponding values of $A,B,C$ are given in
the  last column of Table~1.   This amplitude predicts  the distributions,
(\ref{eq:cosf}) and (\ref{eq:cosmu}).  The $\cos \theta_{\pi}^*$
distribution for P3 shown in Fig.~\ref{figthree} (b) ~differs  a lot from
that for P0 in Fig.~\ref{figone} (a), and
gets much closer to the observed
data.    The lesson from this example is that  once we adopt
(\ref{eq:ampfin}) as the amplitude  for $\upss$, we predict (i) the flat
$\cos\theta_f$  distribution, (ii) $( 1 + \cos^{2} \theta_{l} )$
distribution for
the polar angle of a muon, independent of actual forms of $f_{S}, f_{D}$
and $\delta_{0} (q^{2})$.  This is not the case for $\cos \theta_{\pi}^*$
distribution and thus cannot  be reliably calculated unless they are known.
The actual functional forms of $f_{S}, f_{D}$ and $\delta_{0} (q^{2})$ can
be extracted from the measurement of the joint distribution, $d^{2}\Gamma
/ dm_{\pi\pi} d\cos\theta_{\pi}^{*}$, as one can derive from
(\ref{eq:dcrosect}) :
\begin{eqnarray}
&&{d^{2} \Gamma \over dm_{\pi\pi}d\cos\theta_{\pi}^*}  \nonumber  \\
& \propto & m_{\pi\pi} | \vec{q} | \beta_{\pi}^{*}~\left[
f_{S}^{2} + f_{D}^{2} \left( \cos^{2} \theta_{\pi}^{*} - {1\over 3}
\right)^{2} + 2 f_{S} f_{D} \cos\delta_{0} \left( \cos^{2} \theta_{\pi}^{*}
- {1\over 3} \right) ~\right]    \\
& = & \left[ C_{0} (q^{2}) + C_{2} (q^{2}) \cos^{2} \theta_{\pi}^{*}
+ C_{4} (q^{2}) \cos^{4} \theta_{\pi}^{*} \right].
\end{eqnarray}
For each $m_{\pi\pi}$ bin, one can measure the $\cos\theta_{\pi}^*$
distribution.  This  determines  $C_{i} (q^{2})$'s, and in turn, three
unknowns, $f_{S}, f_{D}$ and $\delta_{0}(q^{2})$.
In particular,  the decay $\upss$ can  be a
source of the  $S-$wave $\pi\pi$ phase shift  for the  whole elastic region,
$2 m_{\pi} \leq m_{\pi\pi} \leq ( m_{i} - m_{f} ) = 895$ MeV \cite{fuchs}.
This may be  important, since the existing data on the $\pi\pi$ phase shift
between  $m_{K} \leq m_{\pi\pi} \leq 600$ MeV are rather poor in statistics
and one has to make some extrapolation \cite{fuchs1}.

\section{Conclusion}
\label{sec:four}

Concluding,  we reanalyzed the $\upss$ decay   using the most general
matrix element in the lowest order in the pion momentum expansion,
including  the final state interactions of the $\pi \pi$
system in the $I = L = 0$ channel.   The $\pi\pi$ phase shift changes
the $\cos\theta_{\pi}^*$ distributions moderately.  (Compare Figs.~1 (a),
(b) with Figs.~2 (a), (b).)
Compared with the recent data from CLEO, P0 is selected, but
the $\cos\theta_{\pi}^*$ distribution does not agree.
In Sec. \ref{sec:three}, we argued that this distribution is
sensitive to possible higher order corrections in the pion momentum
expansion.  As an illustration, we used a new ansatz for the $\pi\pi$
$S-$wave amplitude, (\ref{eq:fsex}), which is of higher order in the pion
momentum expansion, and satisfies  Adler's condition.  This amplitude
could fit the $m_{\pi\pi}$ spectrum (Fig.~3 (a)).   The resulting
$\cos\theta_{\pi}^*$ distribution (Fig.~3 (b)) is
different from Fig,~1 (a), and gets closer qualitatively
to the measured distribution although not quantitatively.
It would be more proper to do the partial wave analysis
with the $S-$ and $D-$waves, and find out the form factors ($f_{S}
, f_{D}$)  and the phase  shift ($\delta_{0} (q^{2})$) in
(\ref{eq:ampfin})  from the
joint  distributions in $m_{\pi\pi}$ and $\cos \theta_{\pi}^*$ using
(20) and (21).
Since our explanation involves both $S-$ and $D-$wave $\pi\pi$
systems, we may be able to find out the $S-$wave phase shift (or
$\delta_{0} - \delta_{2}$, more precisely) from the decay
$\upss$ by measuring various joint distributions.  Since the available
$m_{\pi\pi}$ is below the $K\bar{K}$ threshold, this decay may provide
information on the phase shift over the whole elastic
region, especially for  $m_{K} \leq m_{\pi\pi} \leq 600$ MeV, where the
current data are rather poor in statistics.

\acknowledgements

We  thank Y. Kubota, G. Moneti, R. Poling, S. Rudaz and M. Voloshin for
discussions on  the subject.
We are grateful to J.L. Rosner for careful reading of the
manuscript and valuable suggestions.
P.K. thanks the Aspen Center for Physics for support and hospitality, where
part of this work was done.
This work was supported by Department of Energy
grant \# DOE--DE--AC02--83ER--40105.
S. C. thanks the Mechanical Engineering Department for financial support.

\begin{figure}
\caption{The $\cos \theta_{\pi}^*$ distributions of $\pi^+$ in the rest
frame of the dipion system with the phase shift included : (a) for P0 and
(b) for P1 (and P2).  $\theta_{\pi}^{*} = 0^{\circ}$ corresponds to
the direction of the dipion system as a whole in the rest frame of
the initial $\Upsilon (3S)$.
The dotted and the dashed curves correspond to the low and the
high $m_{\pi\pi}$ regions, and the solid one represents the sum of the two.}
\label{figone}
\end{figure}

\begin{figure}
\caption{The $\cos \theta_{\pi}^*$ distributions of $\pi^+$ in the rest
frame of the dipion system with the phase shift neglected : (a) for P0 and
(b) for P1 (and P2).  $\theta_{\pi}^{*} = 0^{\circ}$ corresponds to
the direction of the dipion system as a whole in the rest frame of
the initial $\Upsilon (3S)$.
The dotted and the dashed curves correspond to the low and the
high $m_{\pi\pi}$ regions, and the solid one represents the sum of the two.}
\label{figtwo}
\end{figure}

\begin{figure}
\caption{ (a) The $m_{\pi\pi}$ spectrum and (b) the $\cos \theta_{\pi}^*$
distributions for the amplitude, (15), where $f_{S}$ is given by (19)
and  $f_D$ is given by (8) with $D=0$.
For (b), the dotted and the dashed
curves correspond to the low and the high $m_{\pi\pi}$ regions, and
the solid one represents the sum of the two, respectively.}
\label{figthree}
\end{figure}

\nopagebreak

\mediumtext
\begin{table}
\caption{Three sets of parameters for the amplitude (10) and one set for
the amplitude (19)  giving the best $\chi^2$ fit to the
$m_{\pi \pi}$ spectrum.  P0 and P3 correspond to the constrained fit with
$D = 0$.  The parameter $A$ is the overall normalization. \label{table1}}

\begin{tabular}{cccc}
Fits  &   P0  &   P1 ( P2 ) & P3  \\
\tableline
$A$  & $10.5 \pm 0.6$  & $  6.7 \pm 2.1 $ & $163.6 \pm 55.3$ \\
$B$  & $-6.4 \pm 0.7$ & $ -3.7 \pm 3.9 $ & $-0.140 \pm 0.035$  \\
$C$  & $36.2 \pm 6.2$ & $ 7.8 \pm 37.2$ &  $ -0.154 \pm 0.001 $  \\
$D$  &  0.0 (fixed) & $ \mp 1.7 \pm 0.7$  &   0.0 (fixed)  \\
$\chi^2/d.o.f.$ &  12.0 / 7 & 11.1 / 6   & 11.0 / 7 \\
{\rm C.L.}   &  10.2 \%  &  8.4 \%  &   14.0 \%  \\
\end{tabular}
\end{table}

\end{document}